\documentclass[prl,twocolumn]{revtex4}
\usepackage{dcolumn,amsmath}
\usepackage{graphicx}
\usepackage{bm}
\usepackage{hyperref}   

\newcommand{\tref}[1]{Table~\ref{#1}}

\begin{document}
\title{Hyperfine and Zeeman interactions of
the $a(1)[^3\Sigma^+_1]$ state of PbO}

\author{A.N.\ Petrov}\email{alexsandernp@gmail.com}
\affiliation
{Petersburg Nuclear Physics Institute, Gatchina,
             Leningrad district 188300, Russia}

\affiliation{Institute of Physics, Saint Petersburg State University, Saint Petersburg, Petrodvoretz 198904, Russia}

\begin{abstract}
  The role of the interaction with the nearest 
electronic state $^3\Sigma^+_{0^-}$ on the hyperfine structure
and magnetic properties of the $a(1)[^3\Sigma^+_1]$ state of PbO
is assessed. The accounting for this contribution leads to
difference between $g$-factors of the $J=1$ $\Omega$-doublet
levels, $ \Delta g = 37\times10^{-4}$, that is in a good agreement
with the
experimental datum $ \Delta g = 30(8)\times10^{-4}$. 
The contribution of this interaction rapidly grows with $J$.
For $J=30$ the difference of $g$-factors of $\Omega$-doublet states
reaches
100\%; for hyperfine constants it is 18\%.
These differences also depend on
the electric field and for $E=11$~V/cm for $^{207}$PbO the difference 
in $g$-factors turn to zero.
The latter is important for suppressing
systematic effects in the electron electric dipole moment search experiment.
\end{abstract}

\maketitle

The use of $a(1)$ excited state of PbO molecule has been proposed to search for
electric dipole moment (EDM) of the electron \cite{DeMille:00}.
This
experiment is a serious test of the ``new physics'' beyond the Standard Model
including different supersymmetric models \cite{Kozlov:95,
Commins:98,Titov:06amin,Rai08}.
Because of that the molecule was intensively investigated both
theoretically \cite{Kozlov:02, Isaev:04, Petrov:05a, Meyer:06a, Meyer:08} 
and experimentally \cite{Hunter:02, Kawall:04b, Bickman:09}.

 In the adiabatic approximation rotational levels of the $a(1)$ state
of PbO are determined by the effective spin-rotational Hamiltonian

\begin{eqnarray}
\nonumber
 H_{sr}=B' {\bf J}^2 + A_{\parallel} ({\bf J}^e\cdot{\bf n}) (\bf{I}\cdot{\bf n}) + \\
  \mu_{\rm B}G_{\parallel}({\bf J}^e\cdot{\bf n})({\bf B}\cdot{\bf n})   -D {\bf n}\cdot {\bf E}
 \label{Hsr}
\end{eqnarray}
%
Here $\textit{B}'$ is the rotational constant,
{\bf J}, {\bf J}$^e$, {\bf I} are the electron-rotational,
electron and nuclear angular momentum operators, respectively
(in this paper we will measure angular momentum in units of $\hbar$),
{\bf E} and {\bf B} are external electric and magnetic fields,
$D$ is the molecular-frame dipole moment,
${\bf n}$ is a unit vector along the molecular axis, ${\bf \zeta}$, directed from
Pb to O, $\mu_{\rm B}$ is Bohr magneton.
The hyperfine constant $A_{\parallel}$ and $g$-factor
$G_{\parallel}$ are determined by the expressions \cite{Dmitriev:92}

\begin{eqnarray}
   A_{\parallel}=\frac{1}{\Omega} \frac{\mu_{\rm Pb}}{I}
   \langle
   \Psi^e_{^3\Sigma^+_{\pm1}}|\sum_i\left(\frac{\bm{\alpha}_i\times
\bm{r}_i}{r_i^3}\right)
_\zeta|\Psi^e_{^3\Sigma^+_{\pm1}}
   \rangle~,
 \label{All}
\end{eqnarray}

\begin{eqnarray}
   G_{\parallel}=\frac{1}{\Omega}
   \langle
   \Psi^e_{^3\Sigma^+_{\pm1}}|{J}^e_\zeta +  {S}^e_\zeta
|\Psi^e_{^3\Sigma^+_{\pm1}}
   \rangle~,
 \label{Gll}
\end{eqnarray}
 where {\bf S}$^e$ is the electron spin operator, $\mu_{\rm Pb}$ is the
 magnetic moment of $^{207}$Pb, $\bm{\alpha}_i$
 are the Dirac matrices for the $i$-th electron, $\bm{r}_i$ is its
 radius-vector in the coordinate system centered on the Pb atom,
$\Omega= \langle\Psi_{^3\Sigma^+_{\pm1}}| {J}^e_\zeta
|\Psi_{^3\Sigma^+_{\pm1}}\rangle = \pm1 $. From naturally
abundant isotopes only  $^{207}$Pb ($I=1/2$) has nonzero
$\mu_{\rm Pb}$, for $^{208}$Pb and $^{206}$Pb $I=0$ and, therefore,
$\mu_{\rm Pb}=0$.

The parameters $\textit{B}'$, $A_{\parallel}$, $G_{\parallel}$ and $D$
can be obtained both theoretically from calculation of the 
{\it electronic} wavefunction $\Psi^e_{a(1)}$ and
by fitting the experimentally observed transitions
to the parameters of the spin-rotational Hamiltonian (\ref{Hsr}).
Comparison of
theoretical and experimental values gives us information
about accuracy of the calculated wavefunction $\Psi^e_{a(1)}$
and, therefore, also gives information about accuracy of the
calculated effective electric field, $W_d$,
seen by an unpaired electron \cite{Isaev:04, Petrov:05a}. 
Note, that $W_d$ can not be measured independently, but it is required
 for extracting $d_e$ from the EDM experiment. 
The experimentally observed parameters $A_{\parallel}$, $G_{\parallel}$
also can be used for a semiemperical evaluation of $W_d$
\cite{Kozlov:02}.

 Previous investigations of PbO were based on the spin-rotational 
Hamiltonian (\ref{Hsr}). The main goal of the present work is to
account for the interaction with the nearest electronic state
$^3\Sigma^+_{0^-}$, which modifies the form of this Hamiltonian.
To the best of our knowledge this is the first investigation of such kind
for open shell diatomics.

In the present paper the hyperfine structure of rotational levels was obtained
by numerical diagonalization of the Hamiltonian in the basis set
of electronic rotational wavefunctions
\begin{eqnarray} 
\Psi^{e}_{^3\Sigma^+_{\pm1}}\theta^{J}_{M,\pm1}(\alpha,\beta)U_{M_I},
\Psi^{e}_{^3\Sigma^+_{0^-}}\theta^{J}_{M,0}(\alpha,\beta)U_{M_I},
\label{basis}
\end{eqnarray}
where
$\theta^{J}_{M,\Omega}(\alpha,\beta)=\sqrt{(2J+1)/{4\pi}}D^{J}_{M,\Omega}(\alpha,\beta,\gamma=0)$
and $U_{M_I}$ are rotational and nuclear spin wavefunctions,
$M$ and $M_I=\pm1/2$ are projections of the angular
momenta, {\bf J} and {\bf I}, on the laboratory axis $z$.
When { \it electronic} matrix elements are known then matrix elements
on the basis set (\ref{basis}) can be calculated with the help of the
angular momentum algebra \cite{LL77}.
Required diagonal {\it electronic} matrix elements, being, in fact, the
parameters of the spin-rotational Hamiltonian (\ref{Hsr}), are
known from experiments. For the fifth vibrational level of the $a(1)$ state of PbO
they are
$B' = 0.235296 ~{\rm cm}^{-1}$, $A_{\parallel} = -4.1 ~ {\rm GHz}$, $G_{\parallel}
= 1.857$, $D = 1.28 ~{\rm a.u.}$ \cite{Hunter:02, Kawall:04b, Bickman:09}.
For purposes of the present study, it is not required
to account for the small difference between the
rotational constants of $^{206,207,208}$PbO molecules.
The differences in properties discussed below are relevant only to the
fact
that the isotope $^{207}$Pb has hyperfine structure.
The off-diagonal { \it electronic} matrix elements were calculated
in the present study by the configuration interaction method
with the generalized relativistic effective core potential
\cite{Titov:99,Mosyagin:06amin}. The scheme of the calculation is the same
as that in the paper \cite{Petrov:05a}. The calculated matrix elements are

\begin{eqnarray}
    \frac{\Delta}{2}=
	B'\langle\Psi^e_{^3\Sigma^+_{1}}|J^e_+
 |\Psi^e_{^3\Sigma^+_{0}}
   \rangle~ = 0.17 ~{\rm cm}^{-1},
 \label{double}
\end{eqnarray}

\begin{eqnarray}
   \frac{\mu_{\rm Pb}}{I}
   \langle
   \Psi^e_{^3\Sigma^+_{1}}|\sum_i\left(\frac{\bm{\alpha}_i\times
\bm{r}_i}{r_i^3}\right)
_+|\Psi^e_{^3\Sigma^+_{0}}
   \rangle~ = -0.7 ~{\rm GHz},
 \label{Aperp}
\end{eqnarray}

\begin{eqnarray}
   G_{\perp} = \langle
   \Psi^e_{^3\Sigma^+_{1}}|{J}^e_+ +  {S}^e_+
|\Psi^e_{^3\Sigma^+_{0}} \rangle = 1.5 .
 \label{Gperp}
\end{eqnarray}

It is known that Hamiltonian (\ref{Hsr}) leads to two-fold degeneracy of levels
with
different signs of $\Omega$. 
This degeneracy is in fact only approximate. When the interaction
(\ref{double})
is taken into account each rotational level splits on two sublevels,
called $\Omega$-doublet levels.
One of them is even $(p=1)$ and the other one is odd $(p=-1)$ with respect
to changing the sign of electrons and nuclear coordinates.
In order to reproduce experimental value of the $\Omega$-doubling,
$5.6\,J(J+1)$~MHz \cite{Kawall:04b}, the matrix element (\ref{double}) 
has to be equal to 0.15~cm$^{-1}$. 
We consider this a good agreement, but will use experimental
value hereafter.
The states with $p=(-1)^J$ 
are denoted as $e$ and with $p=(-1)^{J+1}$
as $f$ states. Note that
the 
wavefunctions
$\Psi^{e}_{^3\Sigma^+_{0^-}}\theta^{J}_{M,0}(\alpha,\beta)U_{M_I}$ 
are $f$ states,
and they do not interact (see below) with $e$ states of the $a(1)$,
unless parity is not conserved, due to weak interactions.

Interactions
(\ref{Aperp}) and (\ref{Gperp}) lead to different hyperfine
  structure and magnetic
properties of the $e$ and $f$ levels. One can estimate from the
second order perturbation theory that contribution from the terms
\begin{eqnarray}
\nonumber
|\langle\Psi^{e}_{^3\Sigma^+_{\pm1}}\theta^{J}_{M,\pm1}U_{M_I}
|\hat{H}_{\rm hfs(mag)}|
\Psi^{e}_{^3\Sigma^+_{0^-}}\theta^{J}_{M,0}U_{M_I}\rangle|^2 \\
/ \left( E_{^3\Sigma^+_{1}}-E_{^3\Sigma^+_{0}} \right)
\end{eqnarray}
is small.
Here $\hat{H}_{\rm hfs}$ and $\hat{H}_{\rm mag}$ are Hamiltonians of the 
hyperfine interaction and the interaction with the external magnetic field, respectively. 
However, the terms
\begin{eqnarray}
\nonumber
2{\rm Re}( \langle\Psi^{e}_{^3\Sigma^+_{\pm1}}\theta^{J}_{M,\pm1}U_{M_I}
|2B'{\bf J}{\bf J^e} |
\Psi^{e}_{^3\Sigma^+_{0^-}}\theta^{J}_{M,0}U_{M_I}\rangle \times \\
\nonumber
\langle\Psi^{e}_{^3\Sigma^+_{0^-}}\theta^{J}_{M,0}U_{M_I}
|\hat{H}_{\rm hfs(mag)}|
\Psi^{e}_{^3\Sigma^+_{\pm1}}\theta^{J}_{M,\pm1}U_{M_I}\rangle ) \\
/ \left( {E_{^3\Sigma^+_{1}}-E_{^3\Sigma^+_{0}}} \right)
\label{PT2}
\end{eqnarray}
are much larger and their
influence on the spectrum of the $a(1)$
state is observable. 

In \tref{gfge} we give calculated $g$-factors for $f$ states with
different quantum numbers $J$ of $^{206,208}$PbO molecule. 
For $e$ states calculated $g_e=1.85700$ and is independent on $J$.
We define $g$-factors so that
the Zeeman splitting is equal to $g_{e(f)}\mu_B B_zM/J(J+1)$. Our 
calculations were done using the finite field method.

\begin{table}
\caption{ 
The $g$-factors for $f$-states of $^{206,208}$PbO as a function of $J$. For $e$ states
$g_e=1.85700$ and is independent on $J$.}
\begin{tabular}{cccc}
    J  &   $g_f$    &   J  &  $g_f$ \\
\hline
    1  & 1.86074  &  10  &  2.06255    \\
    2  & 1.86822  &  12  &  2.14848    \\
    3  & 1.87943  &  15  &  2.30537    \\
	4  & 1.89438  &  20  &  2.64142    \\
    6  & 1.93549  &  25  &  3.07055    \\
    8  & 1.99155  &  30  &  3.59256    \\
\hline
\end{tabular}
\label{gfge}
\end{table}
The obtained difference $g_f-g_e = 37\times10^{-4}$ for $J=1$
is in good agreement with the experimental result 
$g_f-g_e = 30(8)\times10^{-4}$ \cite{Kawall:04b}. As it is seen
from \tref{gfge} the difference is rapidly increasing with $J$, and for
$J=30$ $g_f$ is about two times larger than $g_e$.
Another point to note is that matrix elements (\ref{double}) and (\ref{Gperp}) do not
contribute
to $g_e$ and it remains $J$-independent and unchanged. This is due to the
mentioned above parity selection rule. 
Limiting by the terms (\ref{PT2}) we obtain 
$$ g_f-g_e = \frac{\Delta \cdot G_{\perp} \cdot J(J+1)}{\left( {E_{^3\Sigma^+_{1}}-E_{^3\Sigma^+_{0}}} \right)}
= 1.87032\times 10^{-3} J(J+1)$$ that is in a good agreement with \tref{gfge}.

In \tref{hfs}, the hyperfine splitting (HFS) calculated
between $F=J-1/2$ and $F=J+1/2$ levels as a function of J is
given for $e$ and $f$ states of $^{207}$PbO. Also the results obtained
by applying Eq. (1) and (2) of ref. \cite{Hunter:02} are listed.
Eqs. (1) and (2) of ref. \cite{Hunter:02} give HFS
in the framework of the Hamiltonian (\ref{Hsr}). The interaction 
with the $^3\Sigma^+_{0^-}$ is not taken into account in the
(\ref{Hsr}), therefore
Eqs. (1) and (2) of ref. \cite{Hunter:02} give the same HFS for $e$ and $f$
states of the $a(1)$.
\begin{table}
\caption{  
Calculated values of HFS as a
function of $J$ for $f$- and $e$-levels of $^{207}$PbO. 
Results of Ref.\cite{Hunter:02} are given in
the last column (see Eqs.(1) and (2) therein). 
These results are for both $f$- and $e$-levels.}
\begin{tabular}{cccc}
    J  &\multicolumn{2}{c}{this work}  & Ref. \cite{Hunter:02} \\
       &  ~~~~f        &   ~~~~e   &    \\
\hline
    1  & 3188    &  3187         &  3195 \\
    2  & 1905    &  1903         &  1913 \\
    3  & 1356    &  1353         &  1358 \\
    5  &  863    &    858        &   859\\
   10  &  458    &     449       &  449 \\
   15  &  317    &     304       &  304 \\
   20  &  248    &     230       &  230 \\
   30  &  181    &     154       &  154 \\
\hline
\end{tabular}
\label{hfs}
\end{table}
Similarly to $g$ factors, the hyperfine structure of $e$
states is not affected when interactions (\ref{double}) and
(\ref{Aperp}) are taken into account. However, there is a small difference
between the hyperfine splittings calculated by Eqs. (1) and (2)
in Ref.\ \cite{Hunter:02} and that calculated for $e$ states
in this paper. This
 difference
is related with the fact that
the mixing between the states with $\Delta J=\pm 1$
in Eqs.\ (1) and (2)
of Ref. \cite{Hunter:02}
is taken into account in the framework of the second order perturbation 
theory, whereas in the current work it is calculated more
accurately, by using the numerical diagonalization of the Hamiltonian.

\begin{figure}
\begin{center}
\includegraphics[scale=0.55]{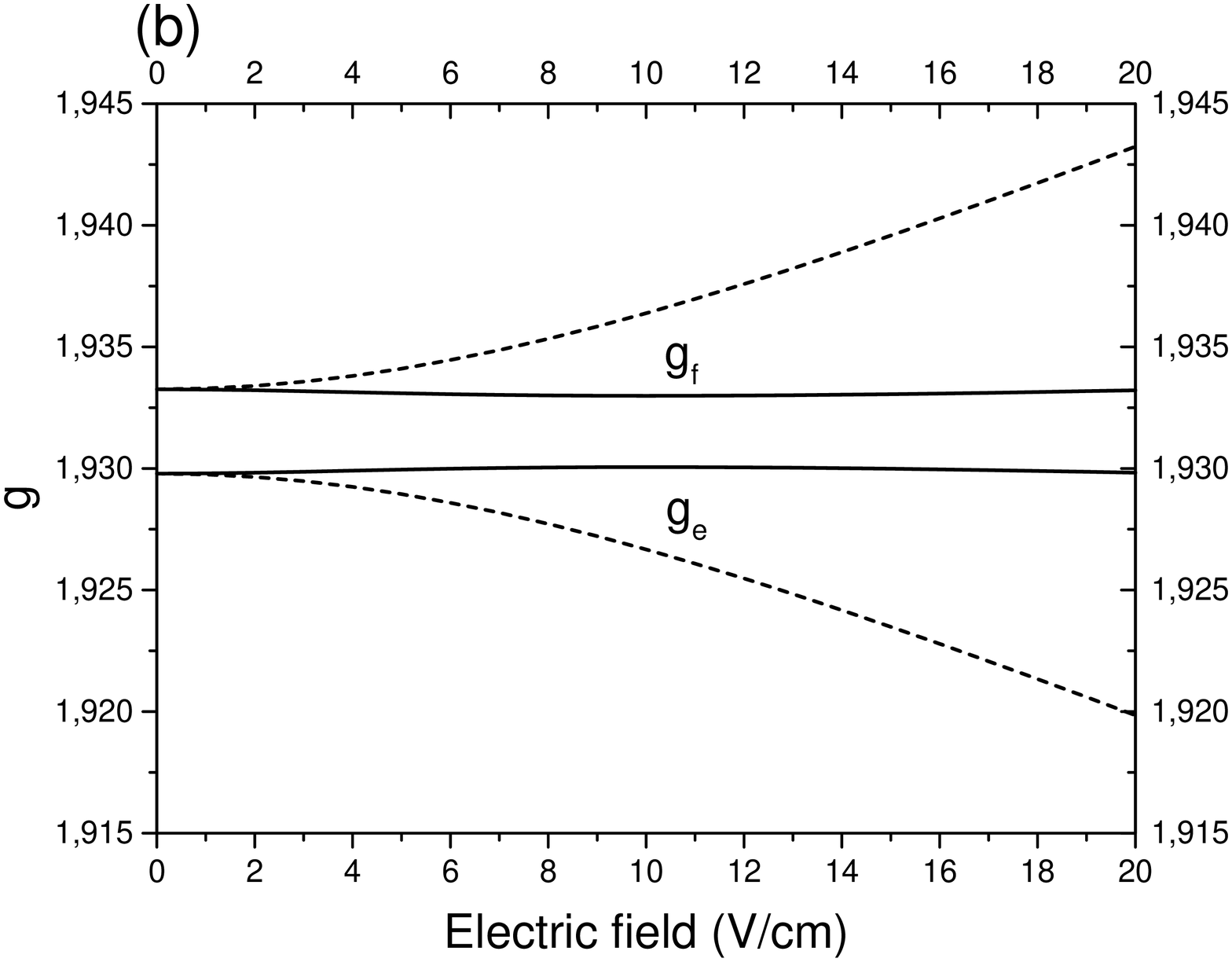}
\includegraphics[scale=0.55]{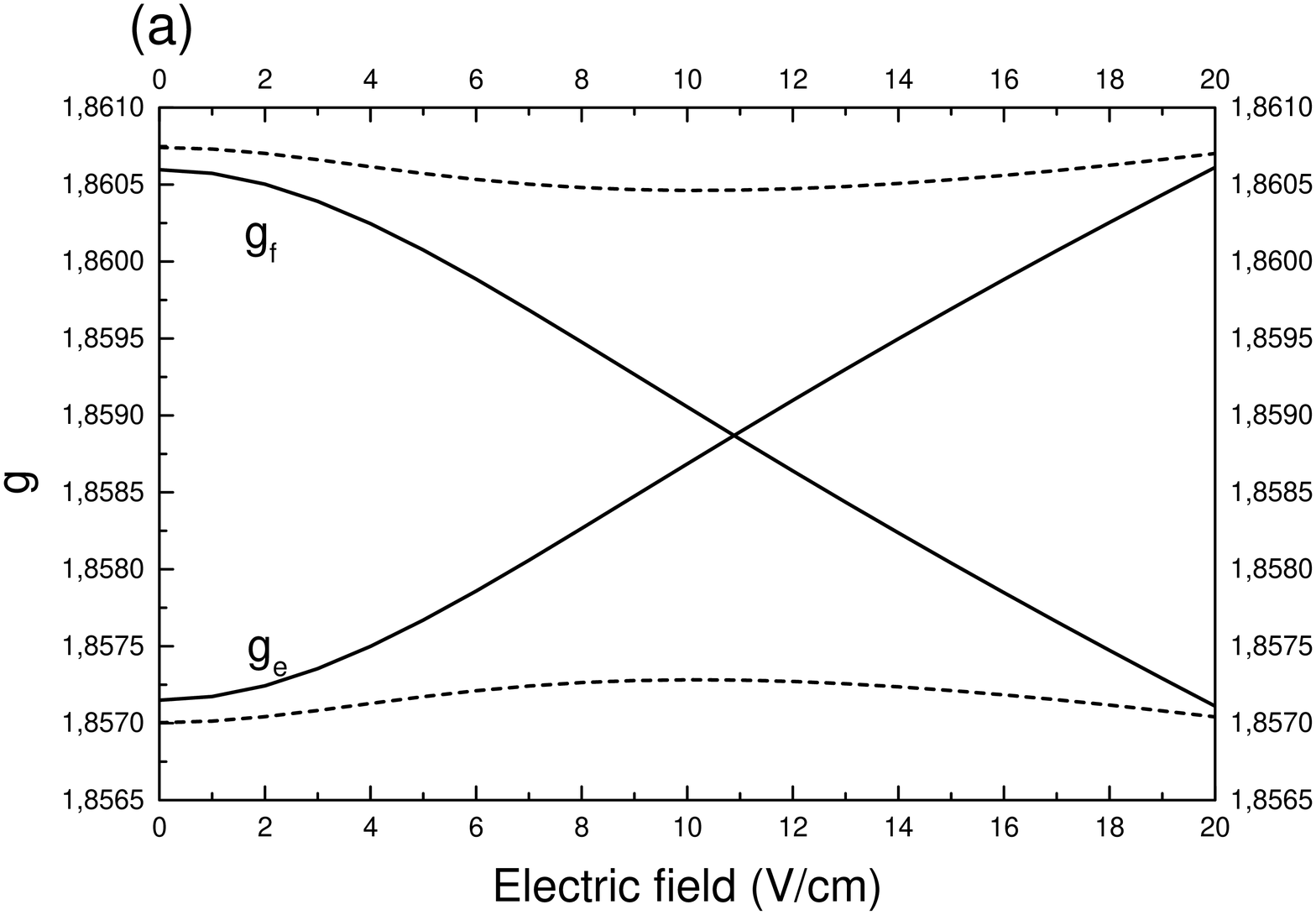}
\caption{\label{gfgecross} Calculated $g$-factors for $e$($g_e$)
and $f$($g_f$) states. (a) Solid lines correspond to $J=1, F=1/2$ 
hyperfine levels of $^{207}$PbO, dashed lines correspond to $J=1$ 
rotational levels of $^{206,208}$PbO. (b) Solid lines correspond to 
$J=1, F=3/2, |M_F|=3/2$, 
dashed lines correspond to $J=1, F=3/2, |M_F|=1/2$ 
hyperfine levels of $^{207}$PbO}
\end{center}
\end{figure}

In the electron EDM search experiment the Stark splitting between
$J=1,M\pm1$ states of the $e$ or $f$ levels is measured.
This Stark effect induced by the interaction with the electron EDM
that violate both parity ($P$) and time reversal ($T$) invariance,
and is not related with the (large) dipole moment $D$ presented
in the (\ref{Hsr}). For details see pp. 1--3 in Ref. \cite{Commins:98}.
In the external electric field the states $J=1$, $M=\pm1$ remain degenerate,
unless both $P$ and $T$ are violated.
However an external magnetic field
remove degeneracy between them and can mimic the existence of the EDM.
For $J=1$ levels the systematics due to spurious magnetic fields can be
suppressed
if the difference between $g_e$ and $g_f$ can be made smaller \cite{Kawall:04b}.
The external electric field mixes $e$ and $f$ levels.
Therefore, on the first glance, one can expect that when increasing
the electric field
the initial small difference between $g_e$ and $g_f$
   can be made
zero. However, it was found in \cite{Kawall:04b} that this
difference for $^{206,208}$PbO is actually increases as the electric field increases.
This fact was explained by M.G.\ Kozlov (see acknowledgments in \cite{Bickman:09})
 by accounting
for the mixing with $J=2$ level. In the present paper 
we reproduce this result for spinless isotopes of led and also
calculate
$g$-factors for $J=1,~ F=1/2,3/2$ states of the $^{207}$PbO.
For $^{207}$PbO, $g$-factors was defined so
that the Zeeman shift is given by 
$$g_{e(f)}\mu_BB_z\frac{F(F+1)+J(J+1)-3/4}{2F(F+1)J(J+1)}M_F.$$
With this definition they will coincide with $g$-factors of
$^{206,208}$PbO in the limit of zero hyperfine interaction.
The corresponding results are
  given
in Fig. (\ref{gfgecross}). 
One can see that difference between $g_e$ and $g_f$ for $J=1, F=3/2$
does not converge to zero as $E$ increases.
However, for $F=1/2, J=1$ at $E \approx 11$~V/cm
$g_e$ and $g_f$ become equal. The plotted $g_e$ and $g_f$ for $J=1$ $^{206,208}$PbO
are in agreement with Fig.~(5) of ref. \cite{Bickman:09}.
Large deviation of $g$-factors for $J=1, F=3/2$ of $^{207}$PbO from
those for $J=1$ of $^{206,208}$PbO is explained by mixing of the
$J=1, F=3/2$ and $J=2, F=3/2$ levels of $^{207}$PbO that is induced by
the hyperfine interaction.

\begin{figure}
\begin{center}
\includegraphics[scale=0.6]{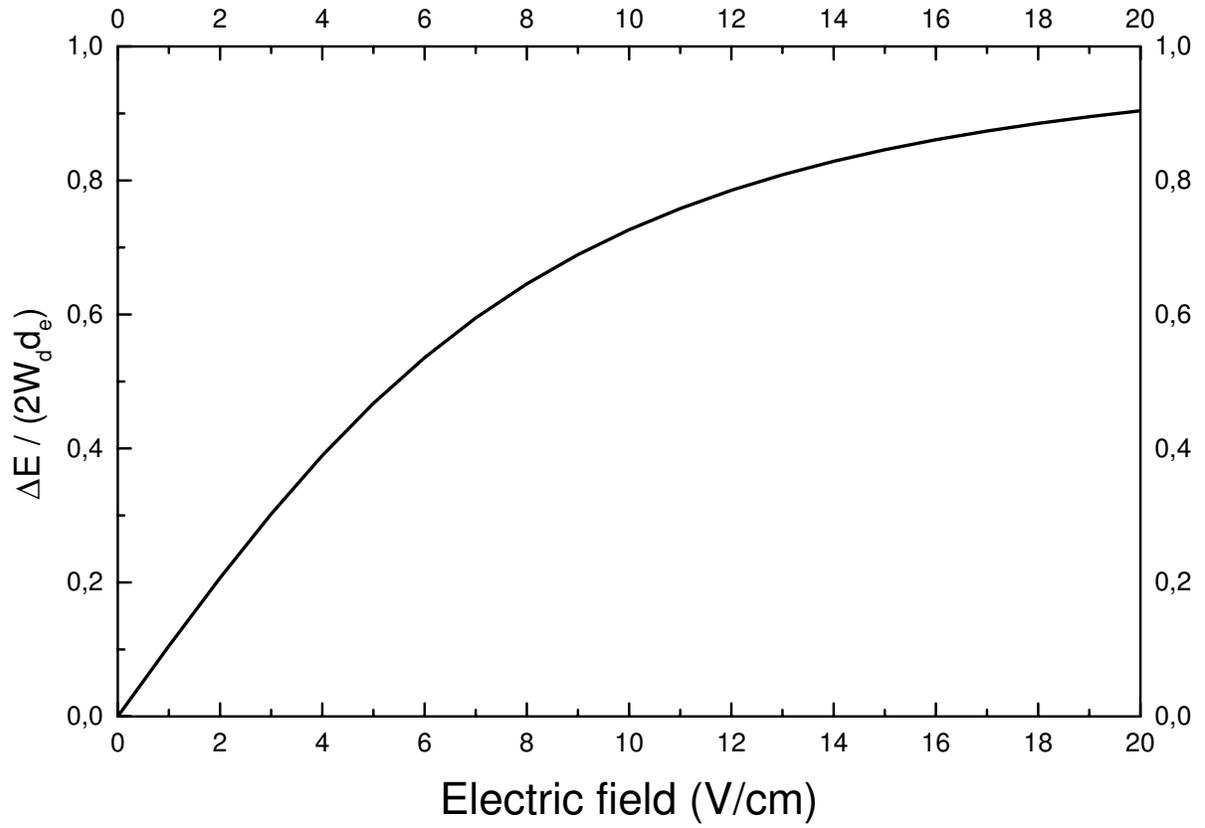}
\caption{\label{edm} 
EDM induced Stark splitting between $M_F=\pm1/2$
levels of $J=1, F=1/2$ state of $^{207}$PbO}
\end{center}
\end{figure}

In the EDM experiment
the maximum Stark splitting,
$2$W$_d$ $\cdot$ d$_e$, between $F=1/2,M_F=\pm1/2$ levels
is achieved for the fully polarized molecule.
In Fig.~(\ref{edm}) we plot
the calculated Stark splitting between $F=1/2,M_F=\pm1/2$ levels
as function of the external electric field. For $E=11$~V/cm the
obtained splitting is about 75\% of the maximal value.

In this work we account for non-adiabatic interaction of $a(1)[^3\Sigma^+_1]$ 
state only with the state $^3\Sigma^+_{0^-}$.
 There are several reasons for this. 
One can see \cite{Huber:79,Polak:93} that the $^3\Sigma^+_{0^-}$ state
is the nearest one to the $a(1)$ state. 
All other states, except $^3\Delta$, are more than an order of magnitude
further away.
Accounting for the non-adiabatic interaction
with the $^3\Delta_1$ state (the same $\Omega=1$ as in $a(1)$)
will lead only to a small
modification of the parameters of the spin-rotational Hamiltonian 
(\ref{Hsr}). Since we use the experimental data, those interactions
with the $^3\Delta_1$ and other $\Omega=1$ states are taken into account.
Though the interaction with $^3\Delta_2$ can not be described in the
framework of the Hamiltonian (\ref{Hsr}),
    it will not lead in the leading order
to the difference in properties of the $f$ and $e$
states that is a topic of this paper.
  Moreover,
our calculation show that the corresponding matrix element
$$
     B'
    \langle\Psi^e_{^3\Sigma^+_{1}}|J^e_-
 |\Psi^e_{^3\Delta^+_2}
   \rangle \approx 3 \times 10^{-3} ~{\rm cm}^{-1}
$$
is small as compared to (\ref{double}). $\Omega=3$ states are not mixed in the leading order
due to the selection rule. The validity of the above approximation is 
approved by the fact that the calculated and the experimentally obtained differences
of the $g$-factors for $e$ and $f$ $J=1$ states
are in good agreement.
%

Finally we have investigated the influence of the interaction with the nearest 
electronic state $^3\Sigma^+_{0^-}$ on the hyperfine structure
and magnetic properties of the $a(1)[^3\Sigma^+_1]$ state. 
We have shown that it is required for its accurate description,
especially for g-factors. One can suppose that similar situation
takes  place also for other diatomics in $\Omega=1$ states.
It is found that the difference between $g_e$ and $g_f$ for $^{207}$PbO is converged to
zero at  $E \approx 11$~V/cm. The latter is important for the suppressing 
systematic effects in the electron EDM search experiment.

I am grateful to M.G. Kozlov and A.V. Titov for
very useful discussions.
This work supported by RFBR Grants  No.~09--03--01034 and
by the Ministry of Education and Science
of Russian Federation (Program for Development of Scientific
Potential of High School) Grant No.~2.1.1/1136


\clearpage


\end{document}